\documentclass[twoside,english]{iopart}
\usepackage[T1]{fontenc}
\usepackage[latin9]{inputenc}
\usepackage{geometry}
\usepackage{babel}
\usepackage{amstext}
\usepackage{amssymb}
\usepackage{graphicx}
\usepackage{esint}
\usepackage[unicode=true,pdfusetitle,
 bookmarks=true,bookmarksnumbered=false,bookmarksopen=false,
 breaklinks=true,pdfborder={0 0 0},backref=false,colorlinks=false]
 {hyperref}

\makeatletter
\usepackage{iopams}
\usepackage{setstack}

\usepackage[numbers, sort&compress]{natbib}

\def\newblock{\hskip .11em plus .33em minus .07em}

\usepackage{iopams}

\makeatother

\begin{document}

\title{Total Internal Reflection of Orbital Angular Momentum Beams }

\author{W. Löffler}

\address{Huygens Laboratory, Leiden University, P.O. Box 9504, 2300 RA Leiden,
The Netherlands}

\ead{loeffler@physics.leidenuniv.nl}

\author{N. Hermosa}

\address{ICFO\textemdash{}Institut de Ciencies Fotoniques, Mediterranean Technology
Park, 08860 Castelldefels (Barcelona), Spain}

\author{Andrea Aiello}

\address{Max Planck Institute for the Science of Light, Günther-Scharowsky-Straße
1/Bldg. 24, 91058 Erlangen, Germany and Institute for Optics, Information
and Photonics, Universität Erlangen-Nürnberg, Staudtstr. 7/B2, 91058
Erlangen}

\author{J. P. Woerdman}

\address{Huygens Laboratory, Leiden University, P.O. Box 9504, 2300 RA Leiden,
The Netherlands}
\begin{abstract}
We investigate how beams with orbital angular momentum (OAM) behave
under total internal reflection. This is studied in two complementary
experiments: In the first experiment, we study geometric shifts of
OAM beams upon total internal reflection (Goos-Hänchen and Imbert-Fedorov
shifts, for each the spatial and angular variant), and in the second
experiment we determine changes in the OAM mode spectrum of a beam,
again upon total internal reflection. As a result we find that in
the first case, the shifts are independent of OAM and beam focussing,
while in the second case, modifications in the OAM spectrum occur
which depend on the input OAM mode as well as on the beam focussing.
This is investigated by experiment and theory. We also show how the
two methods, beam shifts on the one hand, and OAM spectrum changes
on the other, are related theoretically.
\end{abstract}





\maketitle

\section{Introduction}

Simple optical reflection at planar interfaces still offers new surprises.
More than 60 years ago, Goos and Hänchen found that a realistic optical
beam experiences an in-plane displacement with respect to the geometric
optics expected path, under total internal reflection (TIR) \cite{goos1947}.
This is caused by the fact that each of the plane-wave components,
which constitute the beam, picks up a slightly different reflection
coefficient. In other words, even if a beam has a flat wavefront (which
is approximately true for well-collimated beams), beams are not plane
waves and have a finite wave vector spread, which needs to be taken
into account during reflection. Nevertheless, beams are the best approximation
of a geometric-optics ray, and it turns out that beam shifts are the
first-order diffractive corrections to geometric optics. Beam shifts
are polarization dependent. In-plane Goos-Hänchen shifts appear for
$p$ and $s$ linear polarization while the (transverse) Imbert-Fedorov
shifts appear for circular polarization \cite{fedorov1955,imbert1972}.
In recent years, the study of beam shift phenomena got a strong boost
by (amongst other effects) the discovery of the Spin-Hall Effect of
light \cite{bliokhprl2006,hosten2008,aiello2009} and angular shifts
\cite{chan1985,aiellobp2008,merano2009}. In the meantime, shifts
have been discovered not only for light, but also for matter waves
\cite{beenakker2009,haan2010}. 

Beam shifts occurring under partial (external) reflection, such as
angular beam shifts \cite{merano2009}, are known to be sensitive
to the spatial structure of the beam \cite{merano2010}. In particular,
these shifts are sensitive to the beam's orbital angular momentum
\cite{allen1992}, which is appearing, for instance, in Laguerre-Gauss
(LG) laser modes. We have recently shown that one can describe the
diffractive corrections appearing upon reflection also as a change
on the transverse mode spectrum of the beam; notably a pure LG input
mode was found to acquire sidebands \cite{loeffler2012} upon external
reflection. These results were obtained by external reflection of
the input beam. How does this change if we investigate total internal
reflection? Specifically, how does the orbital angular momentum influence
beam shifts, and how is the OAM spectrum modified by TIR? This has
not been studied yet, we shed light on this by theory and experiment.

\section{Shifts of beams with OAM under total internal reflection}

We start with the general case of optical reflection, i.e. we do not
yet specialize to the TIR case. Consider an incoming paraxial, monochromatic
and homogeneously polarized ($\lambda=1,2\equiv p,s$), but otherwise
arbitrary, optical field $\mathbf{U}^{i}(x,y,z)=\sum_{\lambda}U^{i}(x,y,z)a_{\lambda}\hat{\mathbf{x}}_{\lambda}^{i}$
propagating along $\hat{\mathbf{x}}_{3}^{i}$ ($z$ coordinate), where
$\left(a_{1},a_{2}\right)$ is the polarization Jones vector of the
incoming beam. We use dimensionless quantities in units of $1/k_{0}$,
where $k_{0}$ is the wavevector. The coordinate systems and their
unit vectors $\hat{\mathbf{x}}_{\lambda}^{i,r}$ are attached to the
incoming ($i$) and reflected ($r$) beam, respectively. After reflection
at a dielectric interface, the polarization and spatial degree of
freedom are coupled by the Fresnel coefficients $r_{p,s}$ as \cite{merano2010}

\begin{equation}
\mathbf{U}^{r}(x,y,z)=\sum_{\lambda}a_{\lambda}r_{\lambda}U(-x+X_{\lambda},y-Y_{\lambda},z)\hat{\mathbf{x}}_{\lambda}^{r}\equiv\sum_{\lambda}a_{\lambda}r_{\lambda}\mathbf{U}_{\lambda}^{r}.\label{eq:shiftedfield}
\end{equation}

$X_{\lambda}$ and $Y_{\lambda}$ are the polarization-dependent dimensionless
beam shifts:

\numparts\label{eq:shifts}

\begin{equation}
X_{1}=-i\,\partial_{\theta}\left[\ln r_{1}(\theta)\right],\; Y_{1}=i\frac{a_{2}}{a_{1}}\left(1+\frac{r_{2}}{r_{1}}\right)\cot\theta\label{eq:shiftsX}
\end{equation}

\vspace{-5mm}

\begin{equation}
X_{2}=-i\,\partial_{\theta}\left[\ln r_{2}(\theta)\right],\; Y_{2}=-i\frac{a_{1}}{a_{2}}\left(1+\frac{r_{1}}{r_{2}}\right)\cot\theta
\end{equation}
\endnumparts Their real parts yield the spatial beam shifts, and
their imaginary parts the angular beam shifts. They can appear as
longitudinal Goos-Hänchen type shifts \cite{goos1947} $X_{\lambda}$
(along $\hat{x}$ coordinate), or as transverse Imbert-Fedorov type
shifts \cite{imbert1972,bliokhprl2006} $Y_{\lambda}$ along $\hat{y}$.
Transverse shifts $Y_{\lambda}$ require that both $a_{1}$ and $a_{2}$
are finite, such as present in circularly polarized light; this is
not necessary for the longitudinal shifts $X_{\lambda}$. To make
the step from these dimensionless shifts to \emph{observable} shifts,
we have to find the centroid of the reflected beam:

\begin{equation}
\langle\mathbf{R}\rangle(z)=\sum_{\lambda}w_{\lambda}\frac{\int\mathbf{R}\left|U(-x+X_{\lambda},y-Y_{\lambda},z)\right|^{2}\mathrm{d}x\mathrm{d}y}{\int\left|U(-x+X_{\lambda},y-Y_{\lambda},z)\right|^{2}\mathrm{d}x\mathrm{d}y},\label{eq:centroidshift}
\end{equation}
where $w_{\lambda}=\left|r_{\lambda}a_{\lambda}\right|^{2}\big/\sum_{\nu}\left|r_{\nu}a_{\nu}\right|^{2}$
is the fraction of the reflected intensity with polarization $\lambda$,
and $\mathbf{R}=x\hat{\mathbf{x}}_{1}^{r}+y\hat{\mathbf{x}}_{2}^{r}$.
The shift of the centroid depends on the structure of the field, while
the dimensionless beam shifts (Eq.~\ref{eq:shifts}) are independent
of the exact form of $U$. Eq.~\ref{eq:centroidshift} can be calculated
by Taylor expansion around zero shift ($X_{\lambda}=Y_{\lambda}=0$).
 We obtain for the centroid, which is the expectation value for the
2D position vector $\mathbf{R}$ at distance $z$ from the beam waist

\begin{equation}
\langle\mathbf{R}\rangle(z)=\sum_{\lambda}w_{\lambda}\left[\mathrm{Re}\left(\begin{array}{c}
X_{\lambda}\\
Y_{\lambda}
\end{array}\right)+M(z)\,\mathrm{Im}\left(\begin{array}{c}
X_{\lambda}\\
Y_{\lambda}
\end{array}\right)\right],\label{eq:centroidshiftevaled}
\end{equation}
where $M(z)$ is a polarization-independent $2\times2$ matrix, which
depends on the transverse mode of the field \cite{aiello2012distri}.
The $z$-dependent diagonal elements of $M(z)$ describe how angular
shifts influence the apparent position $\langle\mathbf{R}\rangle(z)$,
and the off-diagonal elements effectively mix transverse angular shift
into the longitudinal spatial shift, and the longitudinal angular
shift into the transverse spatial shift. Evaluating Eq.~\ref{eq:centroidshift}
for Laguerre-Gauss beams with $p=0$ and an OAM of $\ell\hbar$, and
using the dimensionless Rayleigh range $\Lambda=2/\theta_{0}^{2}$,
we obtain \cite{merano2010}:

\begin{equation}
\langle\mathbf{R}\rangle(z)=\sum_{\lambda}w_{\lambda}\left[\mathrm{Re}\left(\begin{array}{c}
X_{\lambda}\\
Y_{\lambda}
\end{array}\right)+\left(\begin{array}{cc}
z/\Lambda & -\ell\\
\ell & z/\Lambda
\end{array}\right)\,\mathrm{Im}\left(\begin{array}{c}
X_{\lambda}\\
Y_{\lambda}
\end{array}\right)\right],\label{eq:centroidshiftevaled-1}
\end{equation}
Finally, we specialize to the case of total internal reflection, where
$X_{\lambda}$ and $Y_{\lambda}$ (Eqs.~\ref{eq:shifts}) are real,
which means that mixing via $M(z)$ does not occur; therefore the
shift is expected to be of purely spatial nature and independent of
the orbital angular momentum $\ell$. 

\medskip{}

\begin{figure}
\raggedleft{}\includegraphics[width=8cm]{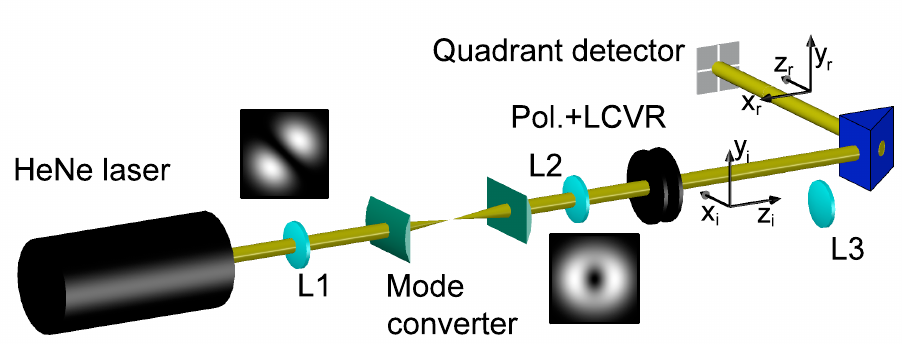}\caption{\label{fig:setupbs}Setup to measure beam shifts under total internal
reflection. A HeNe laser with an intra-cavity wire is set to produce
a clean $HG_{n0}$ mode, and an astigmatic mode converter is used
to convert this mode in a pure Laguerre-Gauss mode. We modulate the
polarization using a variable retarder. The light is then reflected
internally using the prism; and for increasing the beam spread $\theta_{0}$
(focussing), lens L3 can be introduced. The polarization-differential
reflected beam position is determined using a quadrant detector in
combination with a lock-in amplifier. }
\end{figure}

\begin{figure}
\begin{raggedleft}
\includegraphics{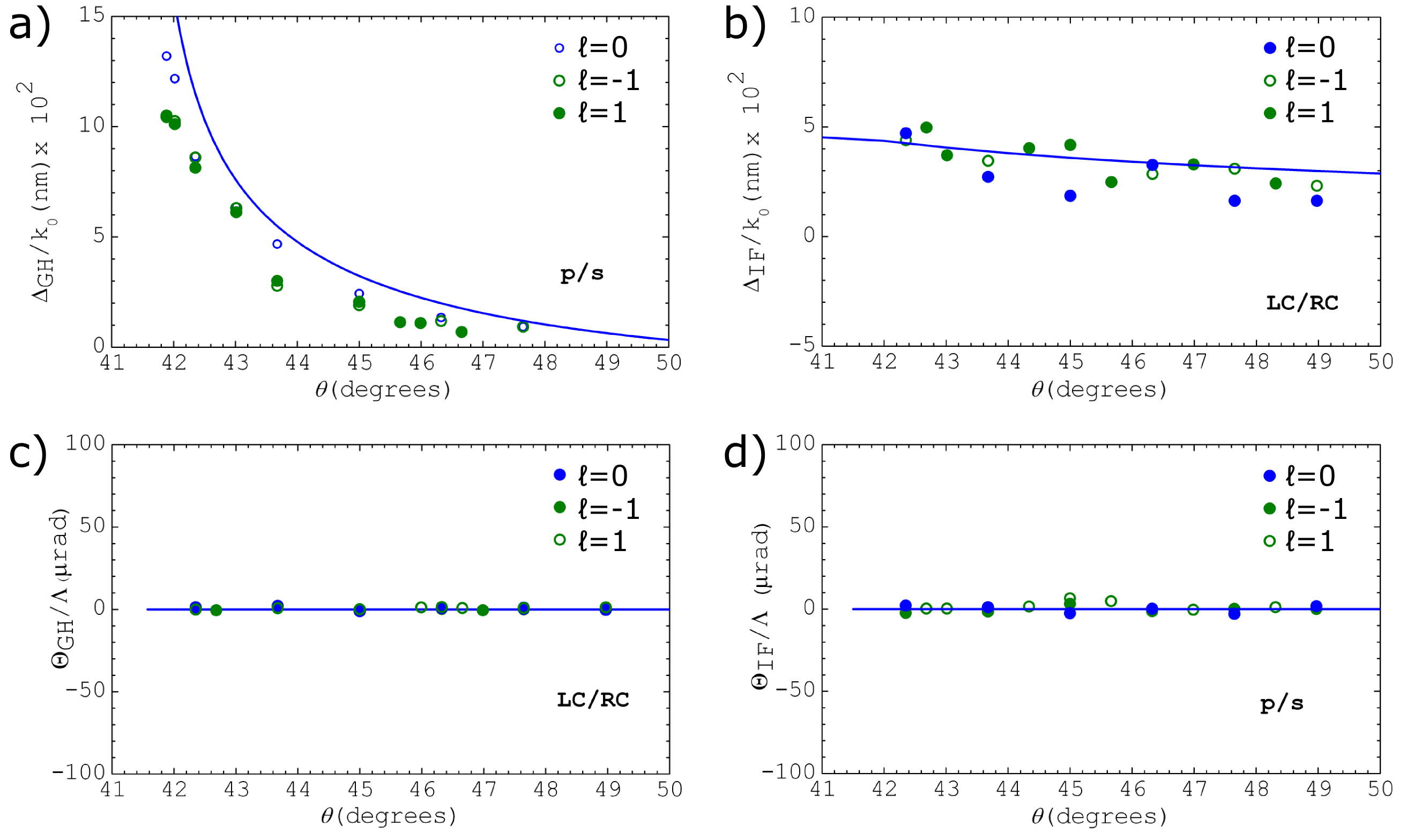}
\par\end{raggedleft}

\raggedleft{}\caption{\label{fig:expbs}Polarization-differential beam shift measurements
for total internal reflection: We see that the spatial Goos-Hänchen
shift $\Delta_{GH}$ (a) and Imbert-Fedorov shift $\Delta_{IF}$ (b)
are present, but independent on the orbital angular momentum $\ell$
of the beam. The angular Goos-Hänchen shift $\Theta_{GH}$ (c) and
angular Imbert-Fedorov shift $\Theta_{IF}$ (d) shifts do not occur.
Shown are the real non-dimensionless quantities as measured in the
lab. }
\end{figure}

This brings us to our first experiment as shown in Fig.~\ref{fig:setupbs},
where we investigate if the OAM mode influences spatial Goos-Hänchen
and Imbert-Fedorov shifts under TIR, and if angular shifts disappear
as expected. The OAM beam is prepared with a custom HeNe laser and
mode conversion. In the laser cavity, a wire (40~$\mathrm{\mu m}$
diameter) is introduced, to enforce a Hermite-Gaussian ($HG_{nm}$)
fundamental mode with $m=0$. This mode is sent through an astigmatic
mode converter \cite{allen1992} consisting of two cylindrical lenses.
The nodal line of the HG mode is oriented at $45^{\circ}$ relative
to the common axis of the mode converter, such that a $HG_{n0}$ mode
is transformed into the $LG_{\ell p}$ mode with $\ell=n$ and $p=0$.
Additional lenses (L1 and L2) are used to ensure mode matching; the
final beam after L2 has a beam waist of 775~$\mu$m. After polarization
modulation with a liquid-crystal variable retarder (LCVR), this beam
is then reflected internally at the hypothenuse of a $45^{\circ}-90^{\circ}-45^{\circ}$
BK7 prism. We measure the polarization-differential beam displacement
using a quadrant detector (which is binned to act effectively as a
split detector), and lock-in techniques (for details, see \cite{merano2010}).
Fig.~\ref{fig:expbs} demonstrates that our theoretical expectations
were accurate: Under TIR, only spatial shifts, here in the polarization
differential form $\Delta_{GH}\equiv\mathrm{Re}\left[X_{1}-X_{2}\right]$
and $\Delta_{IF}\equiv\mathrm{Re}\left[Y_{1}-Y_{2}\right]$ occur,
and are independent of the OAM $\ell$. The angular shifts $\Theta_{GH}\equiv\mathrm{Im}\left[X_{1}-X_{2}\right]$
and $\Theta_{IF}\equiv\mathrm{Im}\left[Y_{1}-Y_{2}\right]$ are identical
to zero. The shifts are also independent of the collimation properties
of the beam as determined by $\theta_{0}$ (data not shown), which
we expect from the discussion above.

\medskip{}

\section{Appearance of OAM sidebands under TIR}

We now come to the second experiment which addresses changes in the
OAM spectrum upon total internal reflection, following \cite{loeffler2012}.
We start by evaluating Eq.~\ref{eq:shiftedfield} now directly. A
first-order Taylor expansion around zero shift ($X_{\lambda}=Y_{\lambda}=0$)
results in 

\begin{equation}
U(-x+X_{\lambda},y-Y_{\lambda},z)\simeq U(-x,y,z)+\mathbf{R}_{\lambda}\cdot\frac{\partial}{\partial\mathbf{R}}U(-x,y,z),\label{eq:taylorser}
\end{equation}
 with $\mathbf{R}=\left(x,y\right)$ and $\mathbf{R}_{\lambda}=\left(X_{\lambda},Y_{\lambda}\right)$.
We substitute $U$ for the well-known normalized Laguerre-Gauss functions,
$U\rightarrow U_{p,\ell}\rightarrow LG_{p}^{\ell}$, where $\ell$
and $p$ are the azimuthal and radial mode indices, respectively.
The spatial Fresnel coefficients $c_{\ell,\ell',p,p'}^{\lambda}$
describe the scattering amplitude for an incoming $LG_{p}^{\ell}$
mode into the $LG_{p'}^{\ell'}$ output channel. These coefficients
are obtained by OAM decomposition of the shifted reflected beam from
Eq.~\ref{eq:taylorser} 
\begin{equation}
c_{\ell,\ell',p,p'}^{\lambda}=\int\text{d}^{2}R\, LG_{p'}^{\ell'\ast}(\mathbf{R})U_{p,\ell}(-x+X_{\lambda},y-Y_{\lambda},z).\label{eq:sfcs}
\end{equation}
If we consider only the OAM part of the Laguerre-Gauss modes (by setting
$p=p'=0$), we find the following simple coefficients (upper and lower
signs refer to the case $\ell\geqslant0$, and $\ell<0$, respectively):

\begin{equation}
c_{\ell,\ell'}^{\lambda}=\left\{ \begin{array}{ll}
\pm Z_{\lambda}^{\pm}\sqrt{|\ell|+1} & \mathrm{for}\;\ensuremath{\ell'=-\ell\mp1}\\
\mp Z_{\lambda}^{\mp}\sqrt{|\ell|} & \mathrm{for}\;\ell'=-\ell\pm1\\
(-1)^{\ell} & \mathrm{for}\;\ell'=-\ell\\
0 & \mathrm{otherwise}
\end{array}\right.\label{eq:sfcsexplicit}
\end{equation}
We see that (in our first-order approximation, see Eq.~\ref{eq:taylorser})
these coefficients couple ``neighboring'' OAM modes with $\ell'=-\ell\pm1$,
where the minus sign stems from image reversal upon reflection. In
other words, the OAM spectrum acquires sidebands upon reflection.
The complex-valued parameters 
\begin{equation}
Z_{\lambda}^{\pm}=\frac{\theta_{0}}{2^{3/2}}(-1)^{\ell}\left(X_{\lambda}\pm i\, Y_{\lambda}\right)\label{eq:z}
\end{equation}
combine all dimensionless shifts. The intensity which appears in a
specific sideband after reflection is $C_{\ell,\ell'}^{\lambda}=|c_{\ell,\ell'}^{\lambda}|^{2}$;
it depends on the strength of the shifts via Eq.~\ref{eq:z}, further
it is proportional to $\ell$ and to the square of the beam opening
angle $\theta_{0}^{2}$. In contrast to the previous case of beam
shifts (Eq.~\ref{eq:centroidshiftevaled-1}), the OAM sideband method
does not discriminate real (spatial) and imaginary (angular) components
of the dimensionless shifts $X_{\lambda}$ and $Y_{\lambda}$, which
is surprising.

\medskip{}

\begin{figure}
\raggedleft{}\includegraphics[width=8cm]{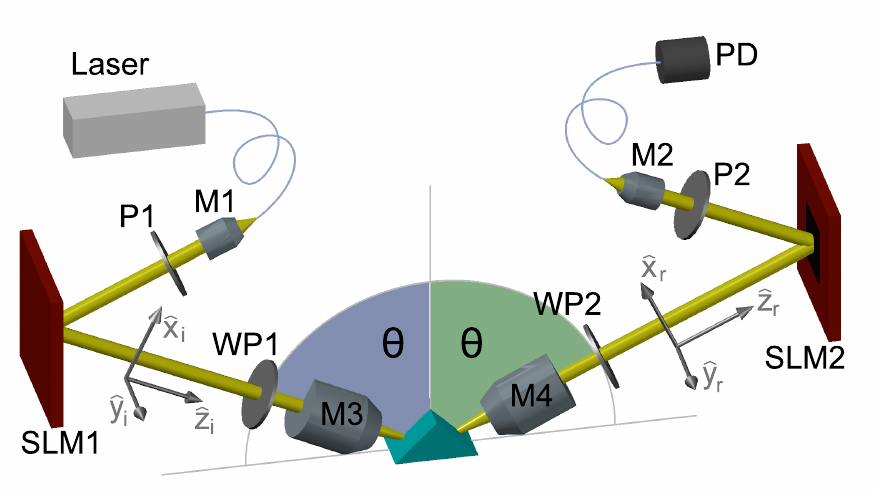}\caption{\label{fig:setupsfc}Experimental setup to measure the spatial Fresnel
coefficients. We prepare the test beam using a fiber collimator (M1),
polarizing optics (P1, WP1), and a spatial light modulator (SLM1).
The beam is reflected internally at the prism hypothenus, and focussing
and re-collimation can be done with M3 and M4 to alter $\theta_{0}$.
The analysis part consists of P2, WP2, SLM2, and M2; the transmitted
intensity is measured using a photo-diode (PD).}
\end{figure}

To demonstrate experimentally the appearance of OAM sidebands under
total internal reflection, we check these dependencies by varying
$\ell$ and $\theta_{0}$. The setup is shown in Fig.~\ref{fig:setupsfc},
where we use again total internal reflection in a $45^{\circ}-90^{\circ}-45^{\circ}$
prism (BK7). We collimate a fiber-coupled 635~nm diode laser with
a $20\times$ objective, then use a spatial light modulator to imprint
the incoming-beam helical phase $\exp(i\ell\phi)$. This method produces,
in terms of Laguerre-Gauss modes, a superposition of modes with the
same azimuthal index $\ell$, but with many radial modes of different
$p$; this superposition depends on the magnitude of $\ell$. This
OAM beam is then reflected internally at the prism hypothenuse. We
analyze the reflected beam by its OAM spectrum with another combination
of SLM and a single-mode fiber, which in turn is connected to a photo
diode. We align the setup for best mode matching between the single
mode fibers of the laser and the detector for the case of $\ell=\ell'=0$.
To vary $\theta_{0}$, we introduce two microscope objectives ($10\times$,
0.25~NA, underfilled aperture). To compensate for small residual
alignment errors, we use polarization modulation by $\lambda/2$ wave
plates on rotation stages, and determine the total polarization-differential
OAM sideband intensity $I_{pd}(\ell)$, see \cite{loeffler2012}.
To obtain a theoretical prediction, we use numerical modeling of the
experiment: This is required because (i) our setup has a transmission
which depends strongly on the selected OAM mode (i.e., transmission
for $\ell'\equiv\ell$, see supplementary information \cite{loeffler2012});
and (ii) the radial-mode superposition as produced by the SLMs has
to be taken into account. In any case, the OAM part of the modes is
well defined throughout our setup.

\begin{figure}
\raggedleft{}\includegraphics{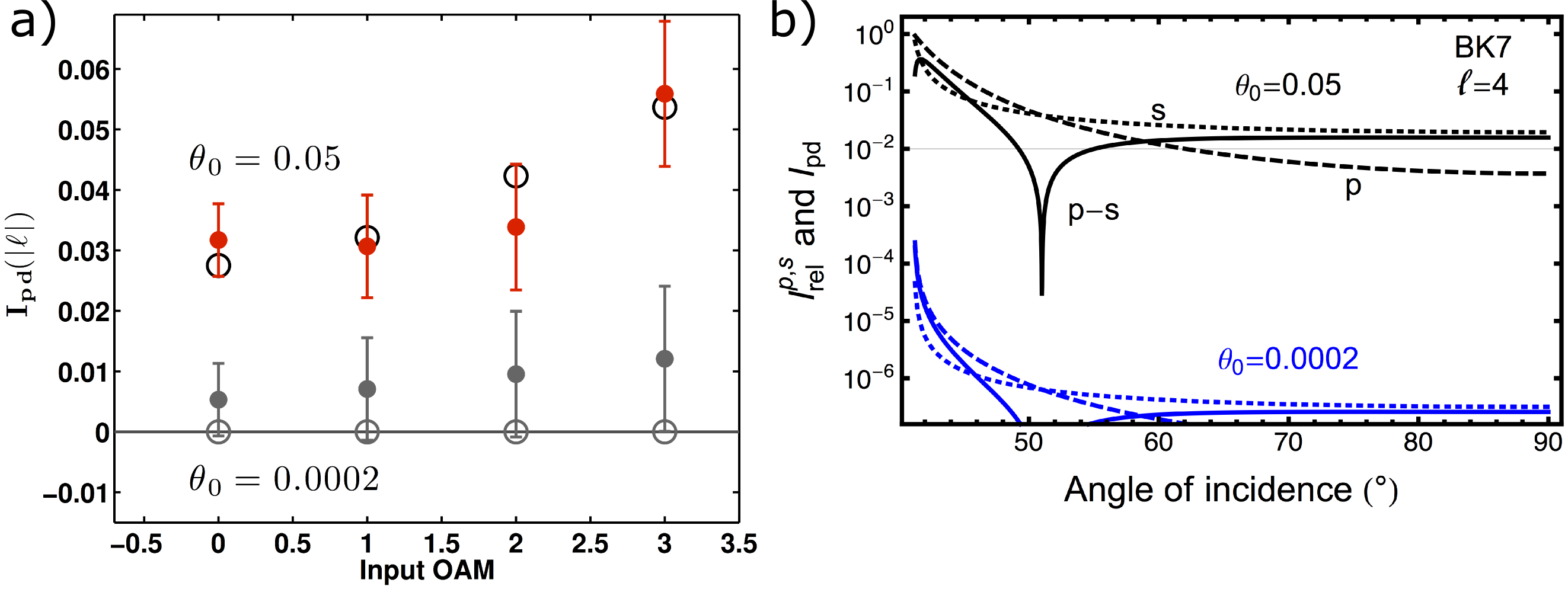}\caption{\label{fig:expsfc}(a): Measurement of the spatial Fresnel coefficients
under total internal reflection. Shown is the polarization-differential
OAM sideband strength $I_{pd}(\ell)$ as a function of the input OAM
$\ell$, for a collimated ($\theta_{0}=0.0002$) and a focussed ($\theta_{0}=0.05)$
beam (filled circles with error bars). The theoretical prediction
(open circles) is based upon numerical simulation of the experiment.
(b): Exact theory for the OAM sideband intensity $I_{pd}(\ell)$ for
$\ell=4$ and the full TIR range of angles of incidence, for true
LG modes with $p=p'=0$. Also shown are the relative sideband intensities
$I_{rel}^{p}$ and $I_{rel}^{s}$, for $p$ and $s$ polarization,
respectively, for both choices of $\theta_{0}$. We conclude that
the OAM sidebands depend on $\ell$ and on $\theta_{0}$; and that
already for mildly focussed beams the OAM sideband strength exceeds
1\% over a large range of angles of incidence.}
\end{figure}

In Fig.~\ref{fig:expsfc}(a) we compare the OAM sideband intensity
from experiment and theory at an angle of $45^{\circ}$, for a collimated
($\theta_{0}=0.0002$) and a focussed ($\theta_{0}=0.05$) beam. We
can confirm that the sideband intensity depends on the input OAM $\ell$
and on the beam opening angle $\theta_{0}$, and agrees with the simulated
data. Fig.~\ref{fig:expsfc}(b) shows the theoretically calculated
total sideband intensity for an incoming $\ell=4$ OAM beam over a
larger range of incident angles. In the case of a focussed beam ($\theta_{0}=0.05$),
the sideband intensity exceeds 1\% for $s$-polarization over the
whole range of total internal reflection. This has to be taken into
account if total internal reflection is used in, e.g., beam steering
applications.

\section{Discussion: OAM beam shifts and sidebands }

We compare now our two experiments: In the first case, we found that
in total internal reflection, beam shifts are independent of the OAM
of the beam, and independent of beam focussing. However, if we measure
the OAM sidebands appearing during total reflection, both properties
influence the result (i.e., the sideband strength). As is obvious
from the similarity of the experiments, both effects describe the
same underlying physical phenomenon: Diffractive corrections to geometric
optics for total internal reflection. We want to explore this relation
briefly theoretically. We start by writing the spatial part (for polarization
$\lambda$) of the reflected field in a quantum-like notation based
on the spatial Fresnel coefficients (we use their full form, see \cite{loeffler2012},
for $\ell>0$):

\begin{eqnarray}
|\mathrm{out}\rangle_{\lambda} & = & \sum_{\ell',p'}a_{\lambda}r_{\lambda}c_{\ell,\ell',p,p'}^{\lambda}|\lambda\rangle|\ell',p'\rangle=a_{\lambda}(-1)^{\ell}|-\ell,p\rangle\\
 &  & +a_{\lambda}Z_{\lambda}^{+}\sqrt{\ell+p+1}|-\ell-1,p\rangle+a_{\lambda}Z_{\lambda}^{+}\sqrt{p}|-\ell-1,p-1\rangle\nonumber \\
 &  & -a_{\lambda}Z_{\lambda}^{-}\sqrt{\ell+p}|-\ell+1,p\rangle-a_{\lambda}Z_{\lambda}^{-}\sqrt{p+1}|-\ell+1,p+1\rangle\nonumber 
\end{eqnarray}
The real space representation of this leads to the centroid as follows: 

\begin{equation}
\langle\mathbf{R}\rangle(z)=\frac{\langle\mathrm{out}|\mathbf{R}|\mathrm{out}\rangle}{\langle\mathrm{out}|\mathrm{out}\rangle}\label{eq:centroidout}
\end{equation}
To be able to compare this to Eq.~\ref{eq:centroidshiftevaled-1},
we need to find the sidebands for pure azimuthal LG input modes with
$p=0$. Eq.~\ref{eq:centroidout} can easily be evaluated by using
the known orthogonality relations of the LG modes, and by recognizing
that the position operator $\mathbf{R}$ can be written as 
\begin{equation}
\mathbf{R}=r\left(\hat{\mathbf{x}}_{1}^{r}\cos\phi+\hat{\mathbf{x}}_{2}^{r}\sin\phi\right)=\frac{r}{2}\left[e^{i\phi}(\hat{\mathbf{x}}_{1}^{r}-i\hat{\mathbf{x}}_{2}^{r})+e^{-i\phi}(\hat{\mathbf{x}}_{1}^{r}+i\hat{\mathbf{x}}_{2}^{r})\right].
\end{equation}
We see that this operator couples modes with $\Delta\ell=\pm1$ in
$\langle\mathrm{out}|\mathbf{R}|\mathrm{out}\rangle$. Straight forward
evaluation of Eq.~\ref{eq:centroidout} leads then exactly to Eq.~\ref{eq:centroidshiftevaled-1},
i.e., the dependency of $\ell$ and $\theta_{0}$ (which was implicit
in the coefficients $c_{\ell,\ell',p,p'}^{\lambda}$) disappears for
total internal reflection. 

Can we give an intuitive explanation why $\theta_{0}$-dependent OAM
sidebands occur for TIR, while angular beam shifts are absent? In
short, the measurement method is different. Spatial beam shifts are
absolute, they do not depend on the beam waist. The OAM spectrum of
a displaced beam, however, depends on the ratio of the displacement
to the beam waist \cite{gibson2004,vasnetsov2005}. If the beam is
collimated, the Goos-Hänchen and Imbert-Fedorov shifts are usually
very small compared to the beam waist, and OAM spectrum is not modified;
if the beam is focussed, appreciable sidebands appear, in agreement
with our experimental results in Fig.~\ref{fig:expsfc}a.

\medskip{}

In conclusion, we have studied the behavior of OAM beams under total
internal reflection. We have investigated this by two complementary
methods: Firstly, by the analysis of optical beam shifts, and secondly
by the observation of modifications in the OAM spectrum of a probe
beam by the spatial Fresnel coefficients. In the first case, we found
that OAM does not modify the spatial beam shifts under total reflection.
In the other case, the opposite is true: The OAM spectrum of a beam
is modified under TIR, and the strength of these modifications increase
with the OAM of the beam as well as its focussing. To resolve this
issue, we have shown how to derive the beam shifts from the spatial
Fresnel coefficients in the short didactical discussion at the top
of this section.

Very recently, a third method to study such diffractive corrections
was found: It turns out that physical reflection induces a split-up
of a high-order vortex into spatially separated first order vortices,
and the splitting is characteristic for a given experimental condition
\cite{dennis2012b}. Based upon our experimental result here, we would
expect that such split-up will occur also under TIR: Vortex splitting
can be explained by coherent background fields \cite{denisenko2008},
and the OAM sidebands, as found here, are of course an example of
such a coherent background field.

\ack{}{}

We acknowledge financial support by NWO and the EU STREP program 255914
(PHORBITECH).

\end{document}